\documentclass{article}
\usepackage[utf8]{inputenc}

\usepackage[numbers]{natbib}
\usepackage{graphicx}
\usepackage{amsmath}
\usepackage{amssymb}
\usepackage{stmaryrd}
\usepackage{subcaption}
\usepackage{float}
\usepackage{url}
\usepackage{listings}
\usepackage{multicol}
\usepackage{hyperref}
\usepackage[noabbrev,capitalise]{cleveref}

\usepackage[frozencache,cachedir=.]{minted}
\usepackage[margin=0.85in]{geometry}
\title{Tensors Fitting Perfectly}
\author{
  Adam Paszke\\Google Research\\\texttt{apaszke@google.com}
  \and
  Brennan Saeta\\Google Research\\\texttt{saeta@google.com}
}
\date{\vspace{-2em}}

\newcommand{\R}{\mathbb{R}}

\newcommand{\call}[2]{\texttt{#1(}#2\texttt{)}}
\newcommand{\ttt}[1]{\texttt{#1}}
\newcommand{\denoteInt}[1]{\mathcal{I}\llbracket#1\rrbracket}
\newcommand{\denoteBool}[1]{\mathcal{B}\llbracket#1\rrbracket}
\newcommand{\denoteShape}[1]{\mathcal{S}\llbracket#1\rrbracket}
\newcommand{\assuming}[1]{~\textbf{assuming}~#1}
\newcommand{\rankof}[1]{\denoteShape{#1}\ttt{.rank}}
\newcommand{\dimension}[2]{\denoteShape{#1}\call{.shape}{#2}}
\newcommand{\loc}{\ttt{loc}}
\newcommand{\hole}{\call{\_\_\_\_}{\loc}}

\newcommand{\lmax}[2]{\textbf{max}\left(#1,~#2\right)}

\begin{document}

\maketitle

\begin{abstract}
Multidimensional arrays (\texttt{NDArray}s) are a central abstraction in modern scientific computing environments.
Unfortunately, they can make reasoning about programs harder as the number of different array shapes used in an execution of a program is usually very large, and they rarely appear explicitly in program text.
To make things worse, many operators make implicit assumptions about the shapes of their inputs: array addition is commonly enriched with broadcasting semantics, while matrix multiplication assumes that the lengths of contracted dimensions are equal.
Because precise reasoning about shapes is crucial to write correct programs using NDArrays, and because shapes are often hard to infer from a quick glance at the program, we developed \emph{Tensors Fitting Perfectly}, a static analysis tool that reasons about \texttt{NDArray} shapes in Swift for TensorFlow programs by synthesizing a set of shape constraints from an abstract interpretation of the program.
It can both (1) check for possible inconsistencies, and (2) provide direct insights about the shapes of intermediate values appearing in the program, including via a mechanism called \emph{shape holes}.
The static analysis works in concert with optional runtime assertions to improve the productivity of program authors.

\end{abstract}

\section{Introduction}

Numerical computing has been completely transformed by the concept of multidimensional arrays and a significant fraction of modern scientific computing workloads are written as compositions of multidimensional array operations, instead of explicit, nested loops operating on scalars.
There are two key reasons for why this abstraction ended up being so useful for efficient execution of scientific simulations.
First, even though most of those programs are written in an imperative language with sequential semantics, most array operators are deeply parallel in nature, and so such programs easily map to modern vectorized hardware which is the main driving force of performance improvements since the end of Dennard scaling.
Second, because the number of array operators invoked is usually orders of magnitude smaller than the number of corresponding scalar operations that need to be performed, the whole computation can be expressed in a relatively inefficient high-level language, enabling rapid development and iteration.
This paradigm, pioneered by APL~\cite{APL}, has fuelled major breakthroughs in science, and led to the development of numerous software packages with similar functionality, such as Matlab or NumPy~\cite{NumPy}.
More recently libraries like Theano~\cite{Theano}, TensorFlow~\cite{TensorFlow}, PyTorch~\cite{PyTorch} and JAX~\cite{JAX} have extended the NDArray abstraction to allow seamless execution of array operators on accelerators like GPUs, and efficiently perform automatic differentiation.

While this paradigm has been very fruitful, one of its biggest unsolved pain points is \textit{shape mismatch errors}.
Every multidimensional array is associated with a finite sequence of natural numbers which represent the domain of indices valid for accessing its elements.
For example, $\R^{n \times m}$~matrices would usually be represented by arrays of shape \texttt{[n, m]} (although the ordering of those dimensions depends on the convention of the libraries and user code).
The length of the shape sequence is often called the \textit{rank}.
If one wants to take a matrix product \texttt{A.dot(B)}, then they need to ensure that both \texttt{A} and \texttt{B} are rank 2, and that the equality \texttt{A.shape[1]~==~B.shape[0]} holds.
In most widely used systems this condition is checked solely at run time, meaning that such errors often go unnoticed until such an operation is invoked.
Because many of these NDArray-based programs run for hours or days before completing (e.g. evaluating model performance on a validation or test dataset, or pre-training on a large unsupervised dataset before fine-tuning for a supervised task), iteratively running the program to debug shape errors can be costly and unproductive.

Scientific computing has greatly benefited from the rise of the open source culture, but unfortunately the difficulty of reasoning about shapes manifests itself especially clearly when analyzing published code.
Most projects treat shapes as irrelevant metadata, and shape contracts of individual user or library functions are left completely implicit.
However, because the semantics of many operations depend on the shapes, having no annotations makes those programs difficult to understand, modify, and maintain, reducing the benefits of free code sharing.
Additionally, when shape mismatches are encountered at runtime, the errors usually originate from the lowest levels of abstraction, often with little context.
In practice, NDArray program authors inevitably need to understand the implementations of the abstractions they consume, reducing the value of the abstraction itself.

One partial solution to that issue, taken by some authors, is to defensively annotate the programs with comments signifying the shape of intermediate variables.
While this makes it much easier for readers to understand what the program is actually doing, such documentation is not checked and can easily get out of date.
Because the shape is just a list of numbers, it seems natural to ask to have it checked by a computer. Unfortunately, no such systems are used today (although some early prototypes are under development~\cite{tsanley}).

Most NDArray libraries support implicit broadcasting, or dynamically increasing the rank of an NDArray by repeating its contents in order to satisfy the shape constraints of a given operation.
For example, adding a vector \texttt{v} of size  $\R^n$ to a matrix \texttt{m} of size $\R^{n \times m}$ results in \texttt{v} being implicitly tiled $m$ times to form a $\R^{n \times m}$ matrix.
Broadcasting improves performance by avoiding materializing copies of \texttt{v} in memory, and implicit broadcasting allows users to write rank-polymorphic code.
Unfortunately, implicit broadcasting makes reasoning about shapes and their resulting errors substantially more difficult and sometimes can even mask them, resulting in a silent divergence from the desired program semantics.

Another possible solution to the shape mismatch problem would be to encode all of the shape constraints in a type system, such that the type checking procedure would additionally prove that no shape errors can occur at run time.
Unfortunately, while type systems that are expressive enough for that exist (e.g. in Haskell \cite{dependentHaskell}), their corresponding type checking problems are undecidable and still require a fair amount of user supervision to have their program accepted, even if it is correct.
On the other hand, type systems used in the most popular imperative languages are not up to the task, and cannot deal with typing of common operations with sophisticated constraints (e.g. functions that would accept inputs of arbitrary ranks, but with some constraints on the trailing dimensions).

\begin{listing}[p]
\centering
\begin{multicols}{2}
\begin{small}
\begin{minted}{swift}
// A sample program entry point to check.
let m = Model()
let batchSize = 12
let input = Tensor(zeros: (12, 32, 32, 3))
let output = m(input)
    |-> [batchSize, 10]

// A simple model used in the program.
//
// A couple shape assertions are
// sprinkled throughout the model.
struct Model {
  let conv = Conv2D(
    filterShape: (2, 2, 3, 5),
    strides: (2, 2))
  let dense = Dense(
    inputSize: ____,
    outputSize: 10)
  
  func callAsFunction(
    _ input: Tensor
  ) -> Tensor {
    let batchSize = input.shape[0]
    let c = relu(conv(input))
    let m = maxPool2D(c, stride: (2, 2))
        |-> [batchSize, 16, 16, 5]  // !!!!
    let dIn = flatten(m)
    let dOut = dense(dIn)
        |-> [batchSize, 10]  // Optional.
    return dOut
  }
}

infix operator |->  // Shape assert operator
func |-> (
    _ a: Tensor,
    _ b: TensorShape
) -> Tensor {
  assert(a.shape == b)
  return a
}

extension Tensor {
  var shape4d: (Int, Int, Int, Int) {
    assert(self.rank == 4)
    let shape = self.shape
    return (
      shape[0], shape[1],
      shape[2], shape[3])
  }
}
\end{minted}
\end{small}

\columnbreak

\begin{small}
\begin{minted}{swift}
// Select operator implementations,
// including shape assertions.

// Computes the valid shape for the output
// of the maxPool2D and conv2D operations.
func validWindowShape(
  _ input: Tensor,
  kernelSize: (Int, Int),
  stride: (Int, Int),
  output: Int
) -> TensorShape {
  let (iN, iH, iW, _) = input.shape4d
  return [
    iN,
    (iH - kernelSize.0) / stride.0 + 1,
    (iW - kernelSize.1) / stride.1 + 1,
    outputs
  ]
}

func maxPool2D(
  _ input: Tensor,
  kernelSize: (Int, Int),
  stride: (Int, Int) = (2, 2)
) -> Tensor {
  let result = TF.maxPool2D(
    input, kernelSize, stride)
  // Shape assertion without shape
  // assertion operator.
  assert(result.shape == validWindowShape(
    input,
    kernelSize: kernelSize,
    stride: stride,
    output: input.shape[3])  // Channels.
  return result
}

func conv2D(
  _ input: Tensor,
  _ weight: Tensor
) -> Tensor {
  let (_, iH, iW, iC) = input.shape4d
  let (kH, kW, iF, oF) = weight.shape4d
  assert(bF == oF); assert(iC == iF)
  assert(iH >= kH); assert(iW >= kW)
  let result = TF.conv2D(input, weight)
      |-> validWindowShape(
            input, kernelSize: (kH, kW),
            stride: (1, 1), output: oF)
  return result
}
\end{minted}
\end{small}

\end{multicols}
\caption{
A sample program instantiating a neural network and evaluating the network at a given input, along with select implementations of library operations and their shape assertions.
The shape assertion operator (\texttt{|->}) is defined in terms of the \texttt{assert} primitive.
\texttt{maxPool2D} and \texttt{conv2D} are examples of sophisticated shape constraints inherent in neural networks.
The static analysis tool highlights a shape error on the line annotated with \texttt{!!!!}; the correct shape for \texttt{m} is: \texttt{[batchSize, 8, 8, 5]}.
Finally, the static analysis tool informs the user that the shape hole (\texttt{\_\_\_\_}) should be replaced with 320.
}
\label{fig:ex}
\end{listing}

In this article, we propose to address the problem with a combination of a static analysis tool coupled with more precise optional run-time shape checking.
Instead of trying to prove that the user program is correct, our static analysis tool searches for provable failures, and will raise an error if any of those are found.
In this way, while the checking procedure is not complete and can miss some issues, we can provide a low-noise signal.
Run-time assertions complement the static analysis to reason about shapes more accurately, but in case their costs are prohibitive for any application, most programming languages (Swift included) allow skipping their evaluation entirely through a compiler or interpreter flag.

Apart from checking, our tool is also able to provide the user with some hints about the shape semantics of the program through functionality like \textit{shape holes} that will be discussed in Section \ref{sec:shape-holes}.
Future work could integrate functionality into the popular editors to interactively assist the researchers and programmers in their daily tasks.

\section{High level overview}

This article is a description of Tensors Fitting Perfectly, an open-source tool\footnote{https://github.com/google-research/swift-tfp} for finding shape bugs in Swift programs developed as part of the Swift for TensorFlow project~\cite{s4tf}.
It takes a list of Swift files as input and passes those on to the Swift compiler, which is asked to lower it to the Swift Intermediate Language (SIL) instead of producing an executable or a library.
SIL is a static single assignment (SSA) based intermediate representation (IR) of the user program.
One interesting note is that SIL does not make use of phi-instructions, but instead allows each block to take a list of arguments.
All jumps to that block have to specify the values that those arguments are supposed to take.
SIL preserves the full power of the Swift type system, and allows for storing most (e.g. compound) values in virtual registers.
Additionally, while the supported instruction set is quite complex, it already desugars and normalizes most of the numerous language features of Swift into much simpler constructs; for example all closures are already converted into top-level functions at this point.
This step also includes type checking of the program, so it lets us ensure that we only process well-formed code, but unfortunately limits the use in the context of incomplete programs which would be necessary to provide live hints in text editors.

Each top level function in the analyzed SIL file is preprocessed to eliminate loops (Section \ref{sec:loop-elimination}) from its control flow graph (CFG) and to remove the need to analyze memory loads and stores (e.g. stack allocations are hoisted into virtual registers).
Each block appearing in the SIL representation is then subjected to symbolic execution (Section \ref{sec:symbolic-execution}), which attempts to recover the assertions specified by the user, and lift them into the raw form of the constraint language used later.
Initially the representation includes function calls which have to be resolved to produce the canonical form on which checking can be performed.
After processing each function, we produce a \textit{function summary} which includes the set of constraints (including calls to other functions) as well as the expressions corresponding to the arguments and returned values.
Note that the produced summary is not of bounded size.
In general, the inputs to the verification procedure may grow even exponentially together with the size of the program (due to inlining), but the pathological examples generally do not resemble real world programs.

Once an entry point and a path condition is chosen by the verification procedure (Section \ref{sec:checking}), function calls are resolved by recursively instantiating the summaries of called functions in a single constraint system (with appropriate substitutions applied).
The system is then simplified (some operations are expensive to represent in the language of the solver), translated into the solver's intermediate representation (effectively a logical formula), and checked.
If the solver reports that the system is unsatisfiable, a failure is reported to the user.

\section{Implementation details}

In this section, we discuss the inner workings of the tool in more detail.
We start with an overview of all the steps that prepare the program for symbolic execution, and then discuss the algorithm that decides whether it should be accepted or not.
Finally, since the goal is to provide useful insights to our users, we comment on how to present verification failures (i.e. shape errors) in a way that makes them more approachable.

\subsection{Shape Assertions}

The static analysis tool is configured with special knowledge of two symbols: the NDArray type (e.g. Swift for TensorFlow's \texttt{Tensor} type), and the corresponding shape type (e.g. \texttt{TensorShape}).
Standard assertions are sprinkled throughout the implementation of \texttt{Tensor} methods which form the basis for how the tool reasons about the user's code.
Importantly, the tool is not given any special knowledge of the semantics of the operations on NDArray; instead everything is built up from these assertions within the NDArray implementation.

\subsection{Making the control flow graph loop-free}
\label{sec:loop-elimination}

One of the important decisions that have to be made when designing a verification tool based on SMT solvers is how to handle loops or recursion.
Although loops do not play a crucial role in typical NDArray programs, their presence should not prevent the tool from applying the facts derived about the program before the loop to the continuation after the loop finishes.

Because there is no way to express fixed points in standard propositional and first-order logic~\cite{aho1979universality}, there are a number of techniques to model back edges in the control flow graph.
One common approach is to rewrite each loop appearing in the program as a deeply nested tower of conditional statements, effectively assuming that all loops have their trip counts bounded by the unrolling factor.
However, if one considers all trip counts up to the unrolling factor as possible program paths, then this approach would additionally assume that the loops might execute as many times as this threshold, which might cause e.g. some shapes to appear to become negative on those paths if the loop shrinks them.
Because we have tried to err on the side of soundness (i.e. trying to not report errors for programs that can execute successfully) when designing the tool, we did not take that approach.

Instead, we replace each loop with a single conditional statement, representing a choice between running the loop at least once or skipping it entirely.
If the loop executes at least once then, assuming that the loop terminates, we know that there exists the first and the last iteration, and so we repeat the loop body twice.
The only question is what happens with loop carried variables, and the answer is that the first instance uses the regular loop inputs that would get passed in to the first iteration, while the second instance gets ``fresh'' inputs in the sense that they get materialized out of thin air and we do not know of any dependencies between them and any other program variable.
Note that after this transformation, although the program can no longer be executed (as we would need a way to actually supply the values for the second iteration), it can still be analyzed.
What this means is that we effectively reject to analyze how the shapes change throughout the loop, but only want to use the assertions from inside of the loop body to be cross-checked with the ones that appear both after and before the loop.
It also has the benefit of reducing the number of possible program paths that have to be considered.

\begin{listing*}
\begin{multicols}{3}
\begin{verbatim}
// Swift source.

func loopingFn(
    k: Int,
    input: Tensor
) -> Tensor {
  var x = input
  x = preOp(x)
  for i in 0..<k {
    x = loopOp(x)
  }
  return postOp(x)
}
\end{verbatim}

\columnbreak

\begin{verbatim}
// SIL-like SSA repr.

func loopingFn {
bb0(%
  br bb1(loopCount, %
bb1(%
  cond_br %
bb2:
  br bb1(%
bb3:
  return %
}
\end{verbatim}

\columnbreak

\begin{verbatim}
// Function summary.

loopingFn(s1, s7):
  // Before loop
  preOp(s1, s2)
  s4a == s2
  loopOp(s4a, s6a)
  // After loop
  loopOp(s6b, s4b)
  postOp(s4b, s7)
\end{verbatim}
\end{multicols}

\caption{A simple function containing a loop and tensor operations, a SIL-like SSA representation of the function that is used as input to the TFP tool, and a representation of the resulting shape constraint summary inferred from the function's control flow graph. In the function summary, the calls to sub-procedures should be understood to calls to their summaries, which constrain the shapes of input and output tensors.}
\label{fig:loopVars}
\end{listing*}

\subsection{Symbolic execution}
\label{sec:symbolic-execution}

Now that the control flow graph is loop-free the program is executed symbolically to extract the assertions appearing inside it.
We initialize arguments of every block to a variable of their corresponding type, if it is a supported type, and an additional variable is created to represent the value returned from the function.
Note that there are no variables for compound types, so compound arguments are associated with a compound expression containing variables.
SIL does not have a concept of methods (they are lowered to curried top-level functions), so we treat compound data types like \texttt{struct}s as isomorphic to tuples containing their members.

\begin{figure*}[h]
\begin{small}
\begin{align*}
\text{Integer expressions} \quad n ::
&= l \mid v
        && \text{Literals / variables} \\
&\mid \hole
        && \text{Shape hole (\texttt{loc} is a source location)} \\
&\mid \call{rank}{s}
        && \text{Rank (number of dimensions)} \\
&\mid s[c]
        && \text{Dimension size} \\
&\mid n + n \mid n - n \mid n \cdot n \mid n / n
        && \text{Arithmetic (division rounds down)} \\
\text{Boolean expressions} \quad b ::
&= \ttt{true} \mid \ttt{false} \mid v
        && \text{Literals / variables} \\
&\mid \neg b
        && \text{Negation} \\
&\mid b \land \ldots \land b \mid b \lor \ldots \lor b
        && \text{Conjunction / disjunction} \\
&\mid n = n \mid s = s \mid b = b
        && \text{Primitive type equality} \\
&\mid n > n \mid n \geq n \mid n < n \mid n \leq n
        && \text{Relations between naturals} \\
\text{Shape expressions} \quad s ::
&= v
        && \text{Variables} \\
&\mid [n, \ldots, n]
        && \text{Literals} \\
&\mid \call{broadcast}{s, s}
        && \text{Broadcasting operation} \\
\text{Compound expressions} \quad c ::
&= (e, \ldots, e)
        && \text{Tuple} \\
\text{Expressions} \quad e ::
&= \call{integer}{n} \mid \call{boolean}{b} && \\
&\mid \call{shape}{s} \mid \call{compound}{c} 
\end{align*}
\end{small}
\vspace{-2em}
\caption{Language of shape constraints}
\label{fig:constraintLanguage}
\end{figure*}

In the following, a \textit{path condition} for a given simple block should be understood as a logical formula that has to be satisfiable if the block is to be reachable.
Intuitively, when we encounter a program that corresponds to an \texttt{if} statement, the block corresponding to one of its branches will have the checked expression in its path condition, while the other one will have its negation.
This way we ensure that we never attempt to cross-verify assertions made in one branch with those made in the other one, since those can easily be contradictory, while the actual program will only ever be subject to one of them.

We begin the execution by setting the path condition of the function entry block to \texttt{true}, and conditions of all other blocks to \texttt{false}.
When processing a block, all instructions in its body are executed symbolically over a limited set of abstract values which is a superset of the constraint language supported by TFP (Figure \ref{fig:constraintLanguage}) that additionally includes e.g. function pointers and their partial applications.
Any calls to the \texttt{assert} built-in function are added to the set of constraints, guarded by the block's path condition.
Once a block-terminating instruction is reached, the set of constraints is extended with equations between the symbolic values of the terminator operands and the arguments of the successor blocks (if the terminator is a jump), or with the variable representing the function result.
Finally, the successor blocks have their path condition extended by taking a disjunction of their current condition and a conjunction of the current block's path condition with the condition derived from the terminator if possible (e.g. if it is a conditional jump) or \texttt{true} otherwise (e.g. if it is an unconditional jump).

\subsection{Representing shape formulas in UFNIA}

To verify specifications represented in the constraint language (Figure \ref{fig:constraintLanguage}), we translate those into formulas over the UFNIA logic (uninterpreted functions and non-linear integer arithmetic).
All shape variables are represented as uninterpreted functions from integers to integers and a single integer encoding its rank.
This representation is motivated by the fact that we cannot constrain the domain to a finite subset of natural numbers a priori, because the rank does not have to be known statically.
However, at each point where this function is evaluated, an assertion ensuring that its argument falls between 0 and the rank is added, guaranteeing that the formulas only reason about the fragment of the domain bounded by the corresponding rank variable. The full translation procedure is described in Figure \ref{fig:ufnia}.

Despite the fact that there is no decision procedure for the UFNIA logic and that undecidable problems can easily be encoded in the language of TFP, the Z3 solver we have used in our implementation has not failed to resolve any of the representative examples we have tried so far.
One practical note is that representing shape equality is quite expensive, as it requires the use of universal quantification to express function equality.
It is however one of the most fundamental operations in our constraint language, and because of that we preprocess the constraints before the UFNIA translation by trying to eliminate as many equalities as we can (by eliminating variables that can be substituted without affecting correctness).
With this simple preprocessing step, the verification process is very quick, but one can easily get solver timeouts if it is skipped.

\begin{figure*}[h]
\begin{small}
\begin{align*}
\denoteInt{l} &= l \\
\denoteInt{v} &= v \\
\denoteInt{\hole} &= v_{\loc}~\text{(a variable name unique to program location $\loc$ where the hole appeared)}\\
\denoteInt{\call{rank}{s}} &= \rankof{s} \\
\denoteInt{s[c]} &=
  \begin{cases}
    \dimension{s}{c} \assuming{0 < -c \leq \rankof{s}} & \text{if}~c < 0 \\
    \dimension{s}{\rankof{s} - c} \assuming{0 \leq c < \rankof{s}} & \text{if}~c \geq 0 \\
   \end{cases} \\
\denoteInt{n_1 \odot n_2} &= \denoteInt{n_1} \odot \denoteInt{n_2}~\text{(where $\odot \in \{+,-,\cdot,/\}$)}\\
\\
\denoteBool{\ttt{true}} &= \top \\
\denoteBool{\ttt{false}} &= \bot \\
\denoteBool{v} &= v \\
\denoteBool{\neg b} &= \neg \denoteBool{b} \\
\denoteBool{b_1 \diamond \ldots \diamond b_k} &= \denoteBool{b_1} \diamond \ldots \diamond \denoteBool{b_k}~\text{(where $\diamond \in \{\land, \lor\}$)}  \\
\denoteBool{n_1 = n_2} &= \denoteInt{n_1} = \denoteInt{n_2} \\
\denoteBool{b_1 = b_2} &= \denoteBool{b_1} = \denoteBool{b_2} \\
\denoteBool{n_1 \diamond n_2} &= \denoteInt{n_1} \diamond \denoteInt{n_2}~\text{(where $\diamond \in \{>,\geq,<,\leq\}$)} \\
\denoteBool{s_1 = s2} &= \left(\rankof{s_1} = \rankof{s_2}\right) \land \left(\forall_{i}~\dimension{s_1}{i} = \dimension{s_2}{i}\right) \\
\\
\denoteShape{v} &= (v, v_{\text{rank}}) \\ 
\denoteShape{[n_1, \ldots, n_k]} &= (v_f, v_{f_{\text{rank}}}) \assuming{\rankof{v_f} = k \land \bigwedge \dimension{v_f}{i} = \denoteInt{n_i}}\\
\denoteShape{\call{broadcast}{s_1, s_2}} &= (v_f, v_{f_{\text{rank}}}) \assuming{
   \rankof{v_f} = \lmax{\rankof{s_1}}{\rankof{s_2}} \\
 & \hspace{10.3em} \forall_{i}~\dimension{v_f}{i} = \lmax{\dimension{s_1}{i}}{\dimension{s_2}{i}}} \\
 & \hspace{10.3em} \forall_{i}~\dimension{s_1}{i} = 1 \lor \dimension{s_2}{i} = 1 \\
 & \hspace{21.05em} \lor \dimension{s_1}{i} = \dimension{s_2}{i} \\
\end{align*}
\end{small}
\vspace{-3em}
\caption{Translation of shape constraints into the UFNIA logic. $v_f$ denotes a fresh variable that is created each time a rule is used. Additionally, each time a rule containing $\assuming{\phi}$ is evaluated, the formula $\phi$ is added to the set of assumptions. The result is a conjunction of the output formula and all assumptions made during the translation. The result of $\denoteShape{\cdot}$ is assumed to be a named pair, with first component named \texttt{dims} and second component named \texttt{rank}. Also, $\dimension{x}{i}$ is syntax sugar for $\denoteShape{x}.\texttt{dims}(\rankof{x} - i - 1)$.}
\label{fig:ufnia}
\end{figure*}

\subsection{Checking procedure}
\label{sec:checking}

To verify the program, the algorithm first gathers all possible path conditions appearing in the program constraints.
For each such condition, we first check that this path is actually feasible, i.e. that the blocks labeled by this condition are actually reachable.
This is performed by asserting all constraints \textit{except those that are guarded by the current condition} along with the condition.
The intuition here is that we want to verify whether all the constraints that are in blocks which are less constrained than this one actually allow for its execution.
If the path feasibility check succeeds, we additionally assert the conditions that we have omitted in the check, and run the analysis.

To see why verifying satisfiability of the path condition alone is not sufficient consider the following program:
\begin{minted}{swift}
func g(_ y: Int) {
  if (y == 2) {
    ...
  }
}

func f(_ x: Int) {
  if (x == 1) {
    g(x)
  }
}
\end{minted}
After processing, the system of constraints derived from \texttt{f} would look approximately like this: $(x = 1 \Rightarrow y = x) \land (x = 1 \land y = 2 \Rightarrow true)$. While $x = 1 \land y = 2$ is certainly a satisfiable formula on its own, it is not a good assumption to make in this program given that it implies the path condition of the equality $y = x$.

A careful reader will notice that the above algorithm has a failure mode which is that if blocks that have weaker path conditions contain a shape contradiction then it will be considered as a reason to reject the strongly constrained path as unfeasible.
This however should not be an issue, as it means that the program does in fact contain an error. Because we iterate over all possible path conditions it will be reported when we reach the weakest possible assumption that is necessary to prove it.

Instead of verifying all possible conditions one could also refine the algorithm such that it first considers the most restricted conditions, and only descends to those implied by it if a failure is found or the path is found to be infeasible.
This however is left for a future improvement.

\subsection{Error reporting}

One of the most significant challenges facing verification tools that depend on SMT solvers is providing good error messages.
In our case, it is very hard to analyze the exact cause of failure, and the proofs produced by Z3 on small programs can easily grow to thousands of operations.
Providing those explanations is hard, but making it easy to understand the failures for users is crucial for success of a static analysis tool.
Hence, to give users some insights, we extract an unsat core from the solver (i.e. a subset of the input assertions that lead to a contradiction), and apply a simplification procedure to those.
It looks for expressions of the form \texttt{v~=~...}, and tries to eliminate the variable \texttt{v} from the constraint system by inlining the other side of the equality into other constraints and simplifying those.
As it turns out, this method works quite well in practice.

\begin{listing*}[h!]
\begin{small}
\begin{minted}{swift}
func matmul(_ x: Tensor<Float>, _ y: Tensor<Float>) -> Tensor<Float> {
  assert(x.shape[1] == y.shape[0])
  let r = TensorFlow.matmul(x, y)
  assert(r.shape == [x.shape[0], y.shape[1]])
  return r
}

let x = randn([20, 10])
let y = randn([30, 10])
let z = matmul(x, y)
\end{minted}
\end{small}
\caption{An incorrect use of the matrix multiplication function.}
\label{fig:incorrect-matmul}
\end{listing*}

\begin{listing*}[h!]
\begin{small}
\begin{verbatim}
In main():
Something doesn't fit!
  - 10 = 30
      Asserted at tmp.swift:2
            | func matmul(_ x: Tensor<Float>, _ y: Tensor<Float>) -> Tensor<Float> {
         2  |   assert(x.shape[1] == y.shape[0])
            |   let r = TensorFlow.matmul(x, y)
\end{verbatim}
\end{small}
\caption{An error reported by the TFP tool when checking the program shown in Listing \ref{fig:incorrect-matmul}.}
\label{fig:err}
\end{listing*}

To see what kind of output can be provided with this method, consider a program shown in Listing \ref{fig:incorrect-matmul}.
Running TFP over it, would produce the error in Listing \ref{fig:err}, pointing to a single assertion that (after simplification) is sufficient to prove the error.
Note that it is also printed as \texttt{10 = 30}, which shows what kind of abstract values were deduced for both sides of this equality.

In the future there are two potential improvements that we could consider.
Firstly, instead of using Z3 for all the solving, there's an option of writing a custom model checker that only defers to an SMT solver for arithmetic reasoning, which would allow it to give more context when a failure occurs.
Secondly, one could try to synthesize summaries from Z3 unsat proofs (e.g. match some patterns of most common shape errors), but that might not be robust in practice.

\subsection{Higher level abstractions}

Although our system supports analysing un-annotated user programs, the more assertions appear in the program, the better the error messages produced by TFP become.
Additionally, the tool can know whether the shape error is within the implementation of a function or in the use of the function.
As a result, developing ergonomic syntax for expression assertions is an integral part of a successful tool.

Because the tool derives shape analysis from standard assertions and not from the semantics of the NDArray operations themselves, the entire system gracefully extends to arbitrary user abstractions.
For example, deep learning libraries often vend a variety of \texttt{Layer} abstractions (e.g. a convolutional layer).
Library authors simply add standard \texttt{assert}s within their code.

In \emph{Tensors Fitting Perfectly}, we leverage Swift's support for defining custom operators to define the \texttt{|->} operator which is syntactic sugar for shape assertions.
(See Listing~\ref{fig:ex} for the implementation.)
It asserts the \texttt{Tensor} has a given (symbolic) shape, and returns the \texttt{Tensor} itself, allowing for ergonomic use in variable assignment.
The following is an example in practice:

\begin{small}
\begin{minted}{swift}
func runDigitDetection(
    _ images: Tensor<Float>
) -> Tensor<Float> {
  let batchSize = input.shape[0]
  let output = network(input)
      |-> [batchSize, 10]  // 10 digits
  let predictions = softmax(output)
      |-> [batchSize]
  return predictions
}
\end{minted}
\end{small}

In addition to providing information about shapes statically, these assertions are validated at runtime to help debug shape errors static analysis did not catch.
Note: the runtime checks are disabled when compiling with optimizations to avoid introducing any overheads for high performance applications.

\subsection{Shape holes}
\label{sec:shape-holes}

Finding shape errors can definitely make writing correct programs easier, but that functionality alone does not really improve one's understanding of the shape semantics of the code.
Ideally, one could imagine having an editing environment in which e.g. hovering over a tensor expression or variable would reveal its shape in relation to other variables present in scope, or some globally defined constants (like batch size for example).
Such a tight integration is not yet supported by our tool, but we do provide a way to synthesize constants in the user program based on the shape constraints which we call \textit{shape holes}.

More concretely, whenever an identifier \texttt{\_\_\_\_} (of type \texttt{Int}) is present in the user program, \emph{Tensors Fitting Perfectly} will treat it as a shape hole.
As shown in the constraint language specification (Figure \ref{fig:constraintLanguage}) each hole is associated with its program location.
Later, during the translation to logical formulas (Figure \ref{fig:ufnia}) each hole effectively gets assigned a unique variable labelled by its program location.
Assuming that there are no contradictions in the extracted shape specification, we can retrieve the model from the SMT solver, and see what the valuation of each such special variable is.
Moreover, once we see an example valuation we can query the solver for alternative solutions to determine whether each such value is in fact unique, or to generate a number of examples.
Based on that, if the checking procedure succeeds, TFP will display a message showcasing a few possible values for each program hole, making it very easy to fill e.g. incomplete machine learning model specifications automatically.

For example, consider this program where \texttt{randn} creates a new array with a given size and all entries sampled i.i.d. from the unit Normal distribution and \texttt{matmul} being the matrix multiplication function:

\begin{small}
\begin{minted}{swift}
let x = randn([20, 10])
let y = randn([____, ____])
let z = matmul(x, y)
\end{minted}
\end{small}

Matrix multiplication requires that the contracted dimensions (i.e. second dimension of \texttt{x} and first dimension of \texttt{y}) be equal, while the second dimension of \texttt{y} is completely unconstrained in this case.
Running TFP over it would produce the output in Listing \ref{fig:holeSoln}.

\begin{listing*}
\begin{small}
\begin{verbatim}
In main() -> ():
  - The hole at tmp.swift:2:19 has to be exactly 10
        |   let x = randn([20, 10])
     2  |   let y = randn([____, ____])
        |   let z = matmul(x, y)
  - Some example values that the hole at tmp.swift:2:25 might take are: 1, 2, 3
        |   let x = randn([20, 10])
     2  |   let y = randn([____, ____])
        |   let z = matmul(x, y)
\end{verbatim}
\end{small}
\caption{
Tool output guiding the user to values for shape holes that allow the program to execute correctly.
There are 2 shape holes---both on line 2---in this program.
The tool prints the valid value when only a single shape value satisfies all the program's constraints (the first shape hole), while suggesting a range of valid values if the constraints do not force a single solution (the second shape hole).
}
\label{fig:holeSoln}
\end{listing*}
This confirms the insights stated in the previous paragraph and indicates that the first dimension of \texttt{y} necessarily has to be of size 10, or a shape error will occur otherwise.
It is clear that this approach can be extended further and used for providing functionality like tooltips for code editors or synthesis of program constants that might be hard to compute by hand (a good example here is e.g. the number of features after flattening of a sequence of convolutional layers).

\section{Related work}

While the approach to shape checking taken in this paper is novel, the technique of applying model checkers for bug finding has been known for a long time.
In fact, our tool works similarly to the sketch described in the seminal work of Engler et al.~\cite{engler2001bugs}: we infer the specification from the user program, and report errors if it is contradictory.
Another tool that has adopted a similar approach is PREfix~\cite{bush2000static}.
However, both of those are mostly focused on finding e.g. memory safety issues instead of reasoning about numerical metadata like shapes.
A good introduction to those methods can be found in an excellent review by the authors of the Z3 SMT solver~\cite{bjorner2014smtVerification}.

Much more academic work has been devoted to the task of proving the absence of errors instead.
If we do not limit ourselves to the shape checking problem stated previously, techniques like abstract interpretation~\cite{cousot2005astree, cousot1998introductionAI} have been quite successful.
However, their reliance on the over-approximation of the set of reachable states means that they can often report errors which are not actually present in the user code.
This does not work very well in the context of shape verification, because one does not want to force all users into a defensive mode where they have to write a lot of code devoted to checking shape conditions.
It has also been known to have trouble scaling to larger programs, but successful reasoning about shapes without annotations crucially depends on extracting knowledge from large fragments of the program via inter-procedural analysis.

Annotation checkers~\cite{jackson1995aspect, flanagan2002extended} also deserve a mention here, but those require numerous extra directives from the user, which is very time-consuming in large programs.
Such a productivity hit certainly outweighs the benefits in many research applications where most programs are run once and then modified or discarded.

When one turns to shape verification, most tools built specifically for that purpose are based on an attempt to design a decidable type system that can effectively prove that no shape errors can occur at run time.
Those are usually explored in the context of functional programming, and some good examples include languages like Remora~\cite{Remora}, Dex~\cite{Dex} or Futhark~\cite{Futhark}.
Unfortunately, many operators used commonly in practice today require features that would make the type system undecidable, meaning that all those approaches cannot possibly scale to the way those programs are expressed today.
This is not to say that the current interfaces are perfect and a decidable dialect cannot be made useful.
It is only that those attempts cannot help the thousands of people writing e.g. NumPy programs today.

Projects utilizing undecidable systems aided by solvers like Hasktorch~\cite{Hasktorch} deserve a mention too.
They can usually assign types to most of the commonly used operators, and the modern solvers are actually powerful enough to discharge most constraints emitted by the type checking procedure.
However, their success critically depends on the interesting programs actually forming a decidable subset of the type checking problem, which can be handled with available solvers.

The approaches discussed above provide a lot of safety and can certainly be beneficial in mission critical applications.
However, many numerical programs are written purely for research purposes, and do not need the full guarantee of correctness, especially at a cost of the verification process restricting the expressiveness of the language or significantly slowing down the development speed.
Additionally, all of the typing approaches have been presented in functional programming languages which are not the tool of choice for the vast majority of users today.
Hence, applying those techniques without significantly disrupting existing workflows seems very hard.
This is why we believe that bug finding approaches like the one proposed in this article are an interesting alternative, as they can provide insights into the programs, while avoiding the need for the user to spend too much time wrangling with the tool.

\section{Conclusion and future work}

This article introduces \emph{Tensors Fitting Perfectly}, a static analysis tool that attempts to infer tensor shape specification from optionally annotated user programs, and find shape errors based on those.
The error checking is based on encoding the specification in the UFNIA logic, and using an SMT solver to answer satisfiability queries.

This is unlike any prior work, where all shape checks are either solely carried out at run time or are encoded as part of the type-checking problem, leading to certain limitations and forcing users to change their workflows significantly.
This work presents a practical alternative that gracefully combines a novel static representation of programs with optional dynamic checks to form a pragmatic and practical system to improve user efficiency.

Apart from that, TFP provides the means to infer values of different shape constants, making it possible for programmers to get interactive feedback about the shapes without executing the code.
All of this has been designed to potentially work with partially incomplete or incorrect code (containing e.g. syntax errors), such that in the future it can be used to power high-level code intelligence tools and provide direct insights from within the editors.
Some other future plans for our work include: (a) Expanding support for more language constructs (e.g. \texttt{var} declarations); (b) Adding support for verification across Swift modules; and (c) Writing front-ends for other languages (e.g. Python).

\bibliographystyle{plain}
\bibliography{references}

\begin{thebibliography}{10}

\bibitem{TensorFlow}
Mart\'{\i}n Abadi, Ashish Agarwal, Paul Barham, Eugene Brevdo, Zhifeng Chen,
  Craig Citro, Greg~S. Corrado, Andy Davis, Jeffrey Dean, Matthieu Devin,
  Sanjay Ghemawat, Ian Goodfellow, Andrew Harp, Geoffrey Irving, Michael Isard,
  Yangqing Jia, Rafal Jozefowicz, Lukasz Kaiser, Manjunath Kudlur, Josh
  Levenberg, Dandelion Man\'{e}, Rajat Monga, Sherry Moore, Derek Murray, Chris
  Olah, Mike Schuster, Jonathon Shlens, Benoit Steiner, Ilya Sutskever, Kunal
  Talwar, Paul Tucker, Vincent Vanhoucke, Vijay Vasudevan, Fernanda Vi\'{e}gas,
  Oriol Vinyals, Pete Warden, Martin Wattenberg, Martin Wicke, Yuan Yu, and
  Xiaoqiang Zheng.
\newblock {TensorFlow}: Large-scale machine learning on heterogeneous systems,
  2015.
\newblock Software available from tensorflow.org.

\bibitem{aho1979universality}
Alfred~V Aho and Jeffrey~D Ullman.
\newblock Universality of data retrieval languages.
\newblock In {\em Proceedings of the 6th ACM SIGACT-SIGPLAN symposium on
  Principles of programming languages}, pages 110--119. ACM, 1979.

\bibitem{bjorner2014smtVerification}
Nikolaj Bj{\o}rner and Leonardo de~Moura.
\newblock Applications of {SMT} solvers to program verification.
\newblock 2014.

\bibitem{JAX}
James Bradbury, Roy Frostig, Peter Hawkins, Matthew~James Johnson, Chris Leary,
  Dougal Maclaurin, and Skye Wanderman-Milne.
\newblock {JAX}: composable transformations of {P}ython+{N}um{P}y programs.
\newblock \url{http://github.com/google/jax}, 2018.

\bibitem{bush2000static}
William~R Bush, Jonathan~D Pincus, and David~J Sielaff.
\newblock A static analyzer for finding dynamic programming errors.
\newblock {\em Software: Practice and Experience}, 30(7):775--802, 2000.

\bibitem{cousot1998introductionAI}
Patrick Cousot and Radhia Cousot.
\newblock Introduction to abstract interpretation.
\newblock 1998.

\bibitem{cousot2005astree}
Patrick Cousot, Radhia Cousot, J{\'e}r{\^o}me Feret, Laurent Mauborgne, Antoine
  Min{\'e}, David Monniaux, and Xavier Rival.
\newblock The astr{\'e}e analyzer.
\newblock In {\em European Symposium on Programming}, pages 21--30. Springer,
  2005.

\bibitem{dependentHaskell}
Richard~A Eisenberg.
\newblock {\em Dependent types in haskell: Theory and practice}.
\newblock PhD thesis, University of Pennsylvania, 2016.

\bibitem{engler2001bugs}
Dawson Engler, David~Yu Chen, Seth Hallem, Andy Chou, and Benjamin Chelf.
\newblock Bugs as deviant behavior: A general approach to inferring errors in
  systems code.
\newblock In {\em ACM SIGOPS Operating Systems Review}, volume~35, pages
  57--72. ACM, 2001.

\bibitem{flanagan2002extended}
Cormac Flanagan, Cormac Flanagan, K~Rustan~M Leino, Mark Lillibridge, Greg
  Nelson, James~B Saxe, and Raymie Stata.
\newblock Extended static checking for java.
\newblock In {\em ACM Sigplan Notices}, volume~37, pages 234--245. ACM, 2002.

\bibitem{Futhark}
Troels Henriksen, Niels~GW Serup, Martin Elsman, Fritz Henglein, and Cosmin~E
  Oancea.
\newblock Futhark: purely functional gpu-programming with nested parallelism
  and in-place array updates.
\newblock In {\em ACM SIGPLAN Notices}, volume~52, pages 556--571. ACM, 2017.

\bibitem{Hasktorch}
Austin Huang, Junjie Hashimoto, Adam Paszke, Sam Stites, and Torsten Scholak.
\newblock Hasktorch.
\newblock \url{https://github.com/hasktorch/hasktorch}, 2019.

\bibitem{APL}
Kenneth~E. Iverson.
\newblock {\em A Programming Language}.
\newblock John Wiley \& Sons, Inc., New York, NY, USA, 1962.

\bibitem{jackson1995aspect}
Daniel Jackson.
\newblock Aspect: Detecting bugs with abstract dependences.
\newblock {\em ACM Transactions on Software Engineering and Methodology
  (TOSEM)}, 4(2):109--145, 1995.

\bibitem{Dex}
Dougal Maclaurin, Radulm Alexey, Matthew J.~Johnson, and Dimitrios Vytiniotis.
\newblock Dex: array programming with typed indices.
\newblock In {\em NeurIPS Program Transformations Workshop}, 2019.

\bibitem{tsanley}
{Nishant Sinha}.
\newblock tsanley.
\newblock \url{https://github.com/ofnote/tsanley}, 2019.

\bibitem{NumPy}
Travis Oliphant.
\newblock {NumPy}: A guide to {NumPy}.
\newblock USA: Trelgol Publishing, 2006.

\bibitem{PyTorch}
Adam Paszke, Sam Gross, Francisco Massa, Adam Lerer, James Bradbury, Gregory
  Chanan, Trevor Killeen, Zeming Lin, Natalia Gimelshein, Luca Antiga, Alban
  Desmaison, Andreas Kopf, Edward Yang, Zachary DeVito, Martin Raison, Alykhan
  Tejani, Sasank Chilamkurthy, Benoit Steiner, Lu~Fang, Junjie Bai, and Soumith
  Chintala.
\newblock Pytorch: An imperative style, high-performance deep learning library.
\newblock In {\em Advances in Neural Information Processing Systems}, 2019.

\bibitem{s4tf}
Brennan Saeta, Denys Shabalin, Marc Rasi, Brad Larson, Xihui Wu, Parker Schuh,
  Michelle Casbon, Daniel Zheng, Saleem Abdulrasool, Aleksandr Efremov, Dave
  Abrahams, Chris Lattner, and Richard Wei.
\newblock {S}wift for {T}ensor{F}low: A portable, flexible platform for deep
  learning.
\newblock {\em MLSys}, 2021.

\bibitem{Remora}
Justin Slepak, Olin Shivers, and Panagiotis Manolios.
\newblock An array-oriented language with static rank polymorphism.
\newblock In {\em European Symposium on Programming Languages and Systems},
  pages 27--46. Springer, 2014.

\bibitem{Theano}
{Theano Development Team}.
\newblock {Theano: A {Python} framework for fast computation of mathematical
  expressions}.
\newblock {\em arXiv e-prints}, abs/1605.02688, May 2016.

\end{thebibliography}
\end{document}